\documentstyle[a4,epsf]{article}

\newcommand{\Lm}{\Lambda}
\newcommand{\Acl}{A_{\rm cl}}
\newcommand{\omg}{\omega}
\newcommand{\sn}{{\rm sn}}
\newcommand{\cn}{{\rm cn}}
\newcommand{\dn}{{\rm dn}}
\newcommand{\gef}{g_{\rm eff}}
\newcommand{\hef}{h_{\rm eff}}
\newcommand{\dms}{\Delta m^{2}}
\newcommand{\Phm}{\Phi_{\rm m}}
\newcommand{\psm}{\psi_{\rm m}}
\newcommand{\lm}{\lambda}
\newcommand{\phcl}{\phi_{\rm cl}}
\newcommand{\chcl}{\chi_{\rm cl}}
\newcommand{\rhng}{\rho_{\rm NG}}
\newcommand{\sgng}{\sigma_{\rm NG}}

\begin{document}

\begin{titlepage}
\null
\begin{flushright}
  {\tt hep-th/0009023}\\
UT-906\\
TIT/HEP-455
\\
September  2000
\end{flushright}

\vskip 2cm
\begin{center}
{\LARGE \bf SUSY Breaking by Overlap of Wave Functions} 
\vskip .5cm
{\LARGE \bf in Coexisting Walls}
\lineskip .75em
\vskip 2.5cm

\normalsize

  {\large \bf Nobuhito Maru~$^{a}$}
\footnote{\it  e-mail address: 
maru@hep-th.phys.s.u-tokyo.ac.jp},  
 {\large \bf 
Norisuke Sakai~$^{b}$}
\footnote{\it  e-mail address: nsakai@th.phys.titech.ac.jp},
 {\large \bf 
Yutaka Sakamura~$^{b}$}
\footnote{\it  e-mail address: sakamura@th.phys.titech.ac.jp} \\
\bigskip
~and~~ {\large \bf 
Ryo Sugisaka~$^{b}$}
\footnote{\it  e-mail address: sugisaka@th.phys.titech.ac.jp}

\vskip 1.5em

{ \it   $^{a}$Department of Physics, University of Tokyo 113-0033, JAPAN \\
and\\
$^{b}$Department of Physics, Tokyo Institute of Technology\\
Tokyo 152-8551, JAPAN  }
\vspace{18mm}

{\bf Abstract}\\[5mm]
{\parbox{13cm}{\hspace{5mm}
SUSY breaking without messenger fields is proposed. 
We assume that our world 
is on a wall and SUSY is broken only by the coexistence of another 
wall with some distance from our wall. 
The Nambu-Goldstone fermion is localized on the distant 
wall. 
Its overlap with the wave functions of physical fields on our wall gives 
the mass splitting of physical fields on our wall 
 thanks to a low-energy theorem. 
We propose that this overlap provides a practical method to evaluate 
mass splitting in models with SUSY breaking due to the coexisting walls. 

}}

\end{center}

\end{titlepage}

\renewcommand{\thefootnote}{\arabic{footnote}}
\baselineskip=0.7cm

\clearpage

{\large\bf Introduction}

\vspace{2mm}

Supersymmetry (SUSY) is one of the most attractive scenarios 
 to solve the hierarchy problem \cite{DGSW}. 
If SUSY is relevant to nature, 
 it must be spontaneously broken at low energies since 
 superparticles have not been observed yet.
Therefore, it is very important to explore 
 mechanisms of SUSY breaking and its mediation to our world.

Recently, much attention is paid to topological objects
 such as D-branes in string theories \cite{D-brane}, 
 BPS domain walls \cite{DW} and junctions \cite{DWJ} 
 in supersymmetric field theories. 
In particular, ``Brane World" scenario \cite{LED,RS}, in which 
 our four-dimensional space-time on these topological objects 
 is embedded in higher dimensional space-time, has opened 
 new directions to flavor physics, cosmology and astrophysics. 
It is well known that 
 these topological objects break supersymmetries partially in general. 
In the light of this fact, 
 it is interesting to study the mechanism of SUSY breaking 
 and its mediation in the context of the brane world scenario. 

The purpose of this paper is to propose 
 a mechanism of SUSY breaking 
 due to the 
coexistence 
of 
walls, 
and to show a practical way to evaluate 
the amount of SUSY breaking for our physical fields 
from the overlap with the wave function of the 
Nambu-Goldstone fermion. 
Our novel point is that the individual wall is 
supersymmetric (a BPS state \cite{BPS}) and 
the SUSY breaking comes about only by the coexistence of these walls 
without any mediating bulk fields (messenger) to communicate SUSY breaking. 
We find that effective SUSY breaking scale observed on our wall becomes 
 exponentially small as the distance between two walls grows, 
whereas the order parameter of the SUSY breaking 
is almost constant. 
We also find that mass splittings become larger for the higher massive modes.  

If a wall is a BPS state, it usually breaks half of the SUSY 
of the original higher dimensional theory. 
Let us suppose that one of such 3-branes is our world and appropriate 
number of supercharges, say four supercharges remain intact. 
In addition to our brane, there may be other branes which preserves 
different combinations of supercharges. 
In such a situation, this ${\cal N}=1$ 
SUSY on our brane is broken only because of 
the existence of the other distant branes. 
This will explain the smallness of the SUSY breaking 
on our brane. 
The simplest situation occurs when we have 
 a wall (we refer to it as ``our wall") 
 and an anti-wall (we refer to it as ``the other wall") 
 parallel to each other. 
Although each wall preserves a half of supersymmetries in the theory, 
orthogonal combinations of supersymmetries are preserved by 
our wall and the other wall. 
Therefore the configuration as a whole 
breaks all of the supersymmetries. 
As we let the other wall going far away, its effect will 
not be felt by observers on our wall and the ${\cal N}=1$ SUSY 
is restored. 
Therefore the smallness of the SUSY breaking can be 
attributed to the distance of the other wall. 
What is crucial for SUSY breaking is the {\em coexistence} of 
 both walls.
In other words, the supersymmetries on our wall is broken by
 the existence of the other wall distant from our wall. 

Note that the SUSY breaking mechanism discussed here involves no 
 complicated SUSY breaking sector on any of the walls. 
We emphasize that we need no bulk field (messenger) to communicate 
 the SUSY breaking in contrast to Refs.\cite{BULK}. 
An idea similar to ours has also been proposed and discussed 
briefly in Ref.\cite{DvaliShifman}. 

\vspace{4mm}

{\large\bf Mass splitting and overlap integral}

\vspace{2mm}

Here we propose a powerful method to find the SUSY breaking mass splitting 
 caused by the coexistence of the other wall.
If the SUSY is broken spontaneously, 
the Nambu-Goldstone fermion $\psi_{\rm NG}$ 
is contained in the supercurrent 
$J^\mu=\sqrt{2} \,i\, f  \gamma^\mu \psi_{\rm NG} + \cdots$ 
with the strength given by 
 the order parameter $f$ of the SUSY breaking, which is 
the square root of the 
energy of the non-trivial background. 
Among the SUSY breaking terms, the low-energy effective Lagrangian contains 
a 
Yukawa coupling 
involving the 
Nambu-Goldstone fermion 
and a scalar $a_{\rm m}$ and spinor $\psi_{\rm m}$ fields 
that belong to the same supermultiplet 
\begin{equation}
 {\cal L}_{\rm yukawa}=\hef a_{\rm m}\psi_{\rm m}\psi_{\rm NG}. \label{NGf}
\end{equation}

The nonlinearly realized 
SUSY manifests itself as a low-energy theorem 
relating 
the Yukawa coupling $\hef$ to the squared mass splitting 
$\dms\equiv m_{B}^{2}-m_{F}^{2}$, where $m_{B}$ and $m_{F}$ are masses of
$a_{\rm m}$ and $\psi_{\rm m}$ respectively 
\cite{lee-wu} 
\begin{equation}
 \hef=-\frac{\dms}{\sqrt{2}f}. \label{GTR}
\end{equation}
The Yukawa coupling $\hef$ is given by an overlap of wave functions of 
$a_{\rm m}$ and $\psi_{\rm m}$ localized on our wall and that of 
 the Nambu-Goldstone fermion $\psi_{\rm NG}$ localized on the other wall. 
However, it is generally difficult to find the exact wave functions 
in the background of two or more walls. 
We will show that we can reliably evaluate 
 an overlap integral of these wave functions 
in the extra dimension 
using an approximation for 
the wave functions in the single-wall background, even if it is difficult 
to find exact modes in the background of two walls. 
Improvement of the approximation is also proposed. 

\vspace{4mm}
 
{\large\bf A model with wall and anti-wall}

\vspace{2mm}

In the following, we consider simple toy models, three-dimensional walls 
 in four-dimensional theory to illustrate 
our mechanism of SUSY breaking without inessential 
 complications. 
Namely we will consider the direction $y=x^2$ as the extra dimension and 
compactify it on $S^{1}$ of radius 
 $R$. 
To test the validity of our approximation methods by 
comparing with the exact result later, 
we will begin with a simple four-dimensional Wess-Zumino model 
with two real positive parameters $g$ and $\Lambda$ 
 in which some wave functions are 
known exactly\footnote
{We follow the convention in Ref.\cite{WessBagger}
} 
\begin{eqnarray}
 &{\cal L}&=\Phi^{\dagger}\Phi |_{\theta^{2}\bar{\theta}^{2}}
  +W(\Phi)|_{\theta^{2}}+{\rm h.c.},  \label{Logn} 
\qquad 
W(\Phi)
=\Lm^{2}\Phi-\frac{g}{3}\Phi^{3}, 
\end{eqnarray}
where $\Phi$ is a chiral superfield 
whose components are defined as 
\begin{displaymath}
 \Phi(x^\mu+i\theta\sigma^\mu\bar\theta,\theta)
=A(x^\mu+i\theta\sigma^\mu\bar\theta)
+\sqrt{2}\theta\psi(x^\mu+i\theta\sigma^\mu\bar\theta)
+\theta^{2}F(x^\mu+i\theta\sigma^\mu\bar\theta).
\end{displaymath}

This model has the following domain wall configuration as a classical 
solution\cite{sakamoto}
\begin{equation}
 \Acl(y)=\frac{k\omg}{g}\sn(\omg(y-y_{0}),k), \;\;\; 
 \omg\equiv\frac{\sqrt{2g}\Lm}{\sqrt{1+k^{2}}}, \label{acl}
\end{equation}
where $\sn(u,k)$ is the Jacobi's elliptic function and $0\leq k\leq 1$. 
The period of this configuration is $4K(k)/\omg$, where $K(k)$ is the 
complete elliptic integral.
Because of the $2\pi R$ periodicity in the extra dimension $y$, 
we take the case of $R=2K(k)/\pi\omg$. 
Then  the solution for $y_{0}=0$ 
represents two walls located at $y=0$ and at $y=\pi R$. 
In the limit of $R\rightarrow\infty$, the wall at $y=0$ becomes 
a BPS domain wall that preserves a half of the SUSY, and  
the other wall at $y=\pi R$ becomes another BPS domain wall 
 preserving the other half of the SUSY. 
The latter may be regarded as an ``anti-(domain) wall''.
Here we will refer to the wall at $y=0$ as ``our wall'' 
which means the 
wall we live on, and will call the wall at $y=\pi R$ as ``the other wall'' 
which gives the source of the small SUSY breaking effect on our wall.

Next we will consider the fluctuation modes around the background 
configuration (\ref{acl}). 
The bosonic mode functions $\phi_{an}(y)$ and $\phi_{bn}(y)$ with eigenvalues 
 $m_{an}^{2}$ and $m_{bn}^{2}$ are defined in terms of the 
differential operators ${\cal O}_{Ba}$ and ${\cal O}_{Bb}$ as 
  \begin{eqnarray}
&\!\!\!&\!\!\!   {\cal O}_{Ba}\equiv 
 -\partial_{y}^{2}-2g(\Lm^{2}-3g\Acl^{2}), \quad
   {\cal O}_{Bb}
\equiv
 -\partial_{y}^{2}+2g(\Lm^{2}+g\Acl^{2}),\nonumber \\
&\!\!\!&\!\!\!   {\cal O}_{Ba}\phi_{an}(y)
= m_{an}^{2} \phi_{an}(y), \quad
   {\cal O}_{Bb} \phi_{bn}(y)
 =
  m_{bn}^{2} \phi_{bn}(y).
  \end{eqnarray}
Assuming the completeness of eigenfunctions $\phi_{an}(y)$ and $\phi_{bn}(y)$, 
 the small real fluctuation fields $a$ and $b$ of the complex scalar field 
$A(x)=\Acl(y)+(a(x)+ib(x))/{\sqrt{2}}$ can be expanded as 
  \begin{eqnarray}
   a(x^{m},y)&\!\!\!=&\!\!\!\sum_{n}\phi_{an}(y)a_{n}(x^{m}), \quad
   b(x^{m},y)
=
\sum_{n}\phi_{bn}(y)b_{n}(x^{m}),\;\;\;\;\;(m=0,1,3).
\label{eq:boson_mode_decomp}
  \end{eqnarray}

  Here $a_{n}(x^{m})$ and $b_{n}(x^{m})$ become scalar fields in 
  three-dimensional effective theory with squared 
masses $m_{an}^{2}$ and $m_{bn}^{2}$
  that are eigenvalues of ${\cal O}_{Ba}$ and ${\cal O}_{Bb}$ respectively.

  Several light modes of $\phi_{an}(y)$ and $\phi_{bn}(y)$ are 
exactly known 
  \begin{eqnarray}
   \phi_{a,-1}(y)
&\!\!\!=&\!\!\!
C_{a,-1}\left\{-\sn^{2}(\omg y,k)+\frac{1+k^{2}+
   \sqrt{1-k^{2}   +k^{4}}}{3k^{2}}\right\}, \nonumber \\
   &\!\!\!&\!\!\!m_{a,-1}^{2}=(1+k^{2}-2\sqrt{1-k^{2}+k^{4}})\omg^{2}, 
\\
   \phi_{a,0}(y)
&\!\!\!=&\!\!\!
C_{a,0}\cn(\omg y,k)\dn(\omg y,k),\;\;\;
   m_{a,0}^{2}=0, \\
   \phi_{a,1}(y)
&\!\!\!=&\!\!\!
C_{a,1}\sn(\omg y,k)\dn(\omg y,k),\;\;\;
   m_{a,1}^{2}=3k^{2}\omg^{2},
  \end{eqnarray}
  where $C_{a,n}$ are positive normalization factors and 
$\cn,\dn$ are the elliptic functions. Other modes are heavier than these.

  Since $m_{a,-1}^{2} < 0$, the field $a_{-1}(x^{m})$ is a tachyon. 
  This instability corresponds to the wall-antiwall annihilation into 
 the vacuum.
  However in the case of the large $R$, the tachyonic squared mass 
  $m_{a,-1}^{2}$ is very close to zero,
  and the system is meta-stable.
  The massless field $a_{0}(x^{m})$ is the Nambu-Goldstone (NG) field 
  corresponding to the breaking of the translational invariance. 

 To expand small fluctuations of fermions into modes, we 
 decompose the Weyl spinor $\psi_{\alpha}$ in four dimensions to two 
real two-component spinors $\psi^{(1)}$ and $\psi^{(2)}$, and 
define a coupled mode equation with an eigenvalue $m_n$ 
\begin{equation}
   \psi_{\alpha}(x)=\frac{1}{\sqrt{2}}(\psi^{(1)}_{\alpha}(x)+
   i\psi^{(2)}_{\alpha}(x)), \quad 
{\cal O}_{\rm F 1}(y) \equiv
\left(\partial_y + 2gA_{\rm cl}(y)\right), \quad 
{\cal O}_{\rm F 2}(y) \equiv 
\left(-\partial_y + 2gA_{\rm cl}(y)\right), 
\label{eq:fermion_op}
\end{equation}
\begin{equation}
{\cal O}_{\rm F 1} \varphi^{(1)}_{n} = -m_n \varphi^{(2)}_{n}, 
\quad 
{\cal O}_{\rm F 2} \varphi^{(2)}_{n}  =-m_n \varphi^{(1)}_{n}.
\label{eq:fermion_eigenvalue_eq}
\end{equation}
The four-dimensional fields $\psi^{(1)}$ and $\psi^{(2)}$ can be expanded 
by these eigenfunctions into three-dimensional fermion fields 
 $\psi^{(1)}_{n}(x^{m})$ and $\psi^{(2)}_{n}(x^{m})$ with mass $m_n$ 
\begin{equation}
   \psi^{(1)}(x^{m},y)
=
\sum_{n}\varphi^{(1)}_{n}(y)\psi^{(1)}_{n}(x^{m}), \quad
   \psi^{(2)}(x^{m},y)=
\sum_{n}\varphi^{(2)}_{n}(y)\psi^{(2)}_{n}(x^{m}). 
\label{eq:fermion_mode_decomp}
\end{equation}
  One can work out explicitly the zero modes 
$m_0=0$ with the normalization factor $C_0$  
  \begin{eqnarray}
   \varphi^{(1)}_{0}(y)&\!\!\!=&\!\!\!
C_{0}\{k\cn(\omg y,k)+\dn(\omg y,k)\}^{2}, \quad 
   \varphi^{(2)}_{0}(y)
=
C_{0}\{k\cn(\omg y,k)-\dn(\omg y,k)\}^{2}.
\label{eq:fermion_zeromode}
  \end{eqnarray}
  We can see that the mode functions $\varphi^{(1)}_{n}(y)$ 
  and $\varphi^{(2)}_{n}(y)$ are localized at $y=0$ and $y=\pi R$ 
  respectively.

  Let us decompose the four-dimensional supercharge 
$
   Q_{\alpha}\equiv(Q^{(1)}_{\alpha}+iQ^{(2)}_{\alpha})/\sqrt{2}
$ 
 into two two-component Majorana supercharges $Q^{(1)}$ and $Q^{(2)}$ 
 which can be regarded as supercharges in three dimensions.
  On our wall $Q^{(2)}$ is broken, but $Q^{(1)}$ is conserved 
in the limit $R\rightarrow \infty$. 
Since $\psi^{(2)}_{0}(x^{m})$ is the Nambu-Goldstone fermion 
corresponding to the broken supercharge $Q^{(1)}$, the massless field 
$\psi^{(2)}_{0}(x^{m})$ is our Nambu-Goldstone fermion $\psi_{\rm NG}(x^{m})$ 
  in Eq.(\ref{NGf}). 
The mode function $\varphi^{(2)}_{0}(y)$ of 
the Nambu-Goldstone fermion is 
 approximately localized on the other wall because 
$Q^{(1)}$ is broken primarily by the presence of 
the other wall. 
  Similarly, the mode function $\varphi^{(1)}_{0}(y)$ is localized 
 on our wall and $\psi^{(1)}(x^m)$ is 
  the Nambu-Goldstone fermion corresponding to $Q^{(2)}$ breaking, 
 whose property as a Nambu-Goldstone fermion is 
  not of our primary concern here. 

 Rewriting the Lagrangian (\ref{Logn}) in terms of the infinite towers of 
the three-dimensional fields in Eqs.(\ref{eq:boson_mode_decomp}) and 
(\ref{eq:fermion_mode_decomp}) 
and integrating out the modes with positive mass squared, 
we obtain a three-dimensional 
low-energy effective Lagrangian for massless and tachyonic fields 
\begin{eqnarray}
 {\cal L}^{(3)}&=&-V_{0}-\frac{1}{2}\partial^{m}a_{-1}\partial_{m}a_{-1}
 -\frac{1}{2}\partial^{m}a_{0}\partial_{m}a_{0}-\frac{i}{2}\psi^{(1)}_{0}
 \partial\hspace{-5.5pt}/\psi^{(1)}_{0}-\frac{i}{2}\psi^{(2)}_{0}
 \partial\hspace{-5.5pt}/\psi^{(2)}_{0} \nonumber \\
 &&-\frac{1}{2}m_{a,-1}^{2}a_{-1}^{2}+\sqrt{2}\gef a_{-1}\psi^{(1)}_{0}
 \psi^{(2)}_{0}, \label{effthry}
\end{eqnarray}
where 
$\partial\hspace{-5.5pt}/\equiv\gamma^{m}_{(3)}\partial_{m}$, 
$\left(\gamma^{m}_{(3)}\right)
\equiv\left(-\sigma^2, i\sigma^3, -i\sigma^1\right)$ 
are $\gamma$-matrices in three dimensions 
and 
\begin{equation}
 V_{0}\equiv \int^{\pi R}_{-\pi R}{\rm d}y\{\Lm^{4}-g^{2}\Acl^{4}(y)\},
\label{eq:vacuum_energy}
\end{equation}
\begin{equation}
 \gef\equiv g\int^{\pi R}_{-\pi R}{\rm d}y \, \phi_{a,-1}\varphi^{(1)}_{0}
 \varphi^{(2)}_{0}
 =gC_{0}^{2}(1-k^{2})^{2}\int^{\pi R}_{-\pi R}{\rm d}y \, \phi_{a,-1}.
 \label{geff}
\end{equation}
The effective Yukawa coupling $\gef$ 
 is proportional to an overlap integral of the mode functions localized on
different walls. 
In Fig.\ref{geff-R}, we show the normalized effective coupling constant 
$\tilde{g}_{\rm eff}\equiv\gef/(g(\sqrt{g}\Lm)^{1 \over 2})$ 
which is  a function of $\sqrt{g} \Lambda R$ only. 
We see that $\gef$ decays exponentially as the wall distance $R$ increases. 

\begin{figure}
 \leavevmode
 \epsfxsize=9cm
 \epsfysize=6cm
 \centerline{\epsfbox{gefR.eps}}
 \caption{The normalized effective coupling $\tilde{g}_{\rm eff}$
 as a function of the distance between the walls.}
 \label{geff-R}
\end{figure}

Since our wall is approximately a BPS domain wall at large $R$,
Eq.(\ref{effthry}) becomes a three-dimensional supersymmetric lagrangian 
in the limit of $R\rightarrow\infty$ with 
$Q^{(1)}$ as the conserved supercharge.
{}From the effective low-energy Lagrangian at finite $R$, we can find 
several terms which show the breaking of $Q^{(1)}$. 
{}Firstly we notice the Yukawa coupling $\gef$ without the associated 
 scalar self-interactions. 
Secondly we see that the tachyonic squared mass $m_{a,-1}^{2}$ can 
also be used as the measure of the SUSY breaking 
since it represents the mass splitting of fermions and bosons. 
We find that the $m_{a,-1}^{2}$ also decay exponentially as $R$ increases, 
and that the ratio of these two quantities $\gef/m_{a,-1}^{2}$ 
becomes constant at large $R$ as shown in Fig.\ref{ratio-R} 
where we used the ``normalized'' tachyonic mass squared 
$\tilde{m}_{a,-1}^{2}\equiv m_{a,-1}^{2}/(g\Lm^{2})$ to make it 
dimensionless.

\begin{figure}
 \leavevmode
 \epsfxsize=9cm
 \epsfysize=6cm
 \centerline{\epsfbox{rtoR.eps}}
 \caption{The ratio of the normalized effective coupling constant 
 $\tilde{g}_{\rm eff}$ and the normalized tachyonic
 mass $\tilde{m}_{a,-1}^{2}$. 
 The dotted line denotes the single-wall approximation,
 and the dashed line denotes the improved single-wall approximation.}
 \label{ratio-R}
\end{figure}

This fact can be understood by noticing that $\gef$ is essentially 
one example of the coupling constant $\hef$ in Eq.(\ref{NGf}).
We define the following scalar mode functions 
\begin{equation}
 \phi^{(1)}_{0}(y)\equiv\frac{1}{\sqrt{2}}(\phi_{-1}(y)+\phi_{0}(y)), 
\quad
 \phi^{(2)}_{0}(y)\equiv\frac{1}{\sqrt{2}}(\phi_{-1}(y)-\phi_{0}(y)),
\label{eq:localized_boson_mode}
\end{equation}
localized on our wall and the other wall respectively. 
By denoting the corresponding fields as  
$ a^{(1)}_{0}(x^{m})\equiv(a_{-1}(x^{m})+a_{0}(x^{m}))/\sqrt{2}$ and  
$ a^{(2)}_{0}(x^{m})\equiv(a_{-1}(x^{m})-a_{0}(x^{m}))/\sqrt{2}$, 
we can rewrite the last term in Eq.(\ref{effthry}) 
involving the effective Yukawa interaction as
\begin{equation}
 {\cal L}^{(3)}_{\rm yukawa}=\gef(a^{(1)}_{0}\psi^{(1)}_{0}\psi^{(2)}_{0}
 +a^{(2)}_{0}\psi^{(2)}_{0}\psi^{(1)}_{0}). \label{l3ykw}
\end{equation}
We can identify ($a^{(1)}_{0}$, $\psi^{(1)}_{0}$) as the supermultiplet 
with respect to approximate $Q^{(1)}$-SUSY and $\psi^{(2)}_{0}$ as 
the Nambu-Goldstone fermion for the broken $Q^{(1)}$. 
Therefore the first term of Eq.(\ref{l3ykw}) precisely gives the desired 
term\footnote{
The remaining second term exhibits the case where our wall and the other wall 
is just interchanged. 
} in Eq.(\ref{NGf}) with $a^{(1)}_{0}$, $\psi^{(1)}_{0}$ and 
$\psi^{(2)}_{0}$ identified as $a_{\rm m}$, $\psi_{\rm m}$ and 
$\psi_{\rm NG}$.
We also rewrite the mass term in Eq.(\ref{effthry}) in terms of
$a^{(1)}_{0}$ and $a^{(2)}_{0}$ and find that the squared mass parameters 
for them are $m_{a,-1}^{2}/2$. 
Thus the low energy theorem in Eq.(\ref{GTR}) 
 now becomes using 
the inverse of the order parameter $f$ of the SUSY breaking as 
\begin{equation}
 \frac{\gef}{m_{a,-1}^{2}}=-\frac{1}{2\sqrt{2}f}. \label{GTR-gef}
\end{equation}

The SUSY algebra implies that the order parameter of the $Q^{(1)}$-SUSY 
breaking $f$ is given by 
the positive square root of the vacuum (classical background) energy density 
$V_0$ in Eq.(\ref{eq:vacuum_energy}) 
\begin{equation}
 f^{2}=\Lm^{4}\int_{-\pi R}^{\pi R}{\rm d}y\left\{1
 -\left(\frac{2k^{2}}{1+k^{2}}\right)^{2}\sn^{4}(\omg y,k)\right\}.
\end{equation}
We find that this order parameter satisfies the low-energy theorem 
(\ref{GTR-gef}) as anticipated and 
becomes 
$ f\rightarrow \frac{4}{\sqrt{3}}\frac{\Lm^{2}}{\sqrt{\sqrt{g}\Lm}}$
in the limit of $R\rightarrow\infty$. 

\vspace{4mm}

{\large\bf Practical method to obtain the mass splitting}

\vspace{2mm}

Next we will consider the situation where some matter fields 
are localized on our wall.
Since $Q^{(1)}$-SUSY is preserved on our wall as $R\rightarrow \infty$, 
matter fields belong to supermultiplets of $Q^{(1)}$-SUSY.
Due to the existence of the other wall, however, $Q^{(1)}$-SUSY is broken 
and mass splitting occurs between a boson and a fermion 
belonging to the same supermultiplet.
As mentioned above, this mass splitting is related to the Yukawa coupling 
$\hef$  and the order parameter $f$ of the 
SUSY breaking by the low-energy theorem in Eq.(\ref{GTR}).
Therefore we can find the mass splitting by calculating the effective 
Yukawa coupling $\hef$. 
This method is powerful, since $\hef$ can be calculated approximately 
for large $R$ by using the mode functions of a single wall 
and thus we can evaluate the mass splitting without solving the difficult 
problem of mass spectra in the background of two coexisting walls.

Let us first test our proposal to obtain the mass splitting 
by an approximate evaluation of the effective Yukawa coupling 
 $\gef$ as an example. 
Since we are interested in the overlap of mode functions of 
supermultiplet of a boson and a 
fermion localized at our wall and that of the Nambu-Goldstone fermion, 
we need a good approximation for the mode functions in the vicinity 
of our wall at $y=0$. 
Therefore we can safely take the supermultiplet mode functions 
in the supersymmetric limit of $R \rightarrow \infty$, namely 
in the single wall limit, 
instead of the exact mode functions. 
The bosonic mode functions defined in Eq.(\ref{eq:localized_boson_mode}) 
and the fermionic ones in Eq.(\ref{eq:fermion_zeromode}) are 
 localized on our wall and becomes in this limit 
\begin{equation} 
 \phi^{(1)}_{0}(y)=\varphi^{(1)}_{0}(y)
=\frac{C} {\cosh^{2}(\sqrt{g}\Lm y)}, 
\qquad -\pi R \le y \le \pi R,
\end{equation}
where $C$ is a normalization factor for the interval $[-\pi R,\pi R]$.

As the first approximation for 
the mode function of the Nambu-Goldstone fermion, 
we are tempted to take the mode function in the presence 
of only one wall at $y=\pi R$ or $y=-\pi R$. 
However, 
to respect the symmetry under $y \rightarrow -y$ and 
the periodic structure, we take a superposition of 
half ($y\le \pi R$) of the mode function 
for a single wall at $y=\pi R$ and another half ($y\ge -\pi R$) 
of the mode function for another single wall 
at $y=- \pi R$. 
We call this a ``single-wall approximation'' 
\begin{equation} 
\varphi^{(2)}_{0}(y) \approx \varphi^{(2)}_{0~{\rm single}}(y) = 
 C\left(\frac{1}{\cosh^{2}(\sqrt{g}\Lm(y-\pi R))}+
 \frac{1}{\cosh^{2}(\sqrt{g}\Lm(y+\pi R))}\right), 
\qquad -\pi R \le y \le \pi R.
\end{equation}
Thus we obtain $\gef$ in Eq.(\ref{geff}) in this approximation as
\begin{equation}
 \gef \approx 2\sqrt{2}g C^{3}\int_{-\pi R}^{\pi R}{\rm d}y\frac{1}
 {\cosh^{4}(\sqrt{g}\Lm y)\cosh^{2}(\sqrt{g}\Lm(y-\pi R))}.
\end{equation}

Since two of three modes involved in the Yukawa coupling have their peaks at
$y=0$, the behavior of the Nambu-Goldstone fermion mode function 
 $\varphi^{(2)}_{0}$ around $y=0$ 
is important to evaluate the above integral.
Note that a fermionic zero mode $\varphi^{(2)}_{0}$ localized at $y=\pi R$ 
can be written \cite{rubakov} as
\begin{equation}
 \varphi^{(2)}_{0}(y)=e^{2g\int^{y}{\rm d}y' \Acl(y')}. \label{rubshp}
\end{equation}
Hence to improve the approximation, we can use the superposition of 
two single walls as a background 
configuration $\Acl$ in Eq.(\ref{rubshp})
\begin{equation}
 \Acl(y)\simeq\frac{\Lm}{\sqrt{g}}\left\{-\tanh(\sqrt{g}\Lm(y+\pi R))
 +\tanh(\sqrt{g}\Lm y)-\tanh(\sqrt{g}\Lm(y-\pi R))\right\}, \quad 
 -\pi R \le y \le \pi R.
 \label{acl_imp}
\end{equation}
We call this as the ``improved'' single-wall approximation and find 
\begin{equation}
\varphi^{(2)}_{0}(y) \approx \varphi^{(2)}_{0~{\rm improved}}(y)
 = C^{(2)}_{0}\frac{\cosh^{2}(\sqrt{g}\Lm y)}
 {\cosh^{2}(\sqrt{g}\Lm(y+\pi R))\cosh^{2}(\sqrt{g}\Lm(y-\pi R))}, 
 \label{ipsgl}
\end{equation}
where $C^{(2)}_{0}$ is a normalization factor for the interval 
$[-\pi R,\pi R]$.
In this improved single-wall approximation we obtain 
\begin{equation}
 \gef \approx \sqrt{2}g C^{2}C^{(2)}_{0}\int_{-\pi R}^{\pi R}{\rm d}y
 \frac{1}{\cosh^{2}(\sqrt{g}\Lm(y+\pi R))\cosh^{2}(\sqrt{g}\Lm y)
 \cosh^{2}(\sqrt{g}\Lm(y-\pi R))}.
\end{equation}

The results of these approximations are shown in 
{}Fig.\ref{ratio-R}. 
We see that the simplest single wall approximation already gives a correct 
order of magnitude estimate for $\gef$ including the sign. 
Therefore it gives a correct information for the mass splitting including 
which is heavier, boson or fermion. 
Moreover, the improved single-wall approximation gives a very 
accurate estimate of the mass splitting, for instance 
within 1~\% in the case of $R\geq 5/(\sqrt{g}R)$. 

To illuminate another aspect of our proposal, we next consider 
a matter chiral superfield 
$
 \Phm =A_{\rm m}+\sqrt{2}\theta\psm + \theta^{2}F_{\rm m}
$
interacting with $\Phi$ 
in the original lagrangian (\ref{Logn}) through an additional superpotential 
$W_{\rm int}=-h\Phi\Phm^{2}$. 
One can easily see that the linearized equations for the matter fields 
are identical to those for the field $\Phi$ building the wall except that 
the coupling $g$ is replaced by $h$. 
By taking the coupling $h$ bigger than $g$, we obtain 
several light modes of $\Phm$ localized on our wall. 
Let us decompose the matter fermion $\psm$ into two real two-component 
spinors $\psi_{{\rm mR}\alpha}$ and 
$\psi_{{\rm mI}\alpha}$ as 
$ \psi_{{\rm m}\alpha}=
 (\psi_{{\rm mR}\alpha}+i\psi_{{\rm mI}\alpha})/\sqrt{2}$. 
Comparing it to Eq.(\ref{eq:fermion_op}), 
we find that the eigenvalue equations for $\psi_{{\rm mR}\alpha}$ and 
$\psi_{{\rm mI}\alpha}$ are obtained by replacing the coupling $g$ by $h$  
in Eq.(\ref{eq:fermion_eigenvalue_eq}) for  $\psi^{(1)}_{\alpha}$ and 
$\psi^{(2)}_{\alpha}$. 
{}For simplicity we consider 
 the limit of $R\rightarrow\infty$ (i.e. single-wall approximation). 
We find that the low-lying eigenvalues are discrete at 
$ m_{{\rm m}n}^{2}= g\Lm^{2}n\left(-n+4h/g\right)$ 
with $n=0,1,2,\cdots < 2h/g$, and that the corresponding 
 eigenfunctions $\rho_{{\rm R}n}(y)$ for the field $\psi_{{\rm mR}\alpha}$ 
are 
\begin{equation}
 \rho_{{\rm R}n}(y)
={N_n \over \left[\cosh(\sqrt{g}\Lm y)\right]^{\frac{2h}{g}-n}}
 {\rm F}\left(-n, 1-n+{4h \over g},1-n+{2h \over g};
 {1-\tanh(\sqrt{g}\Lm y) \over 2}\right),
\label{msvmd}
\end{equation}
where ${\rm F}(\alpha,\beta,\gamma;z)$ is the hypergeometric function 
and $N_n$ is a normalization factor. 
The mode functions for the field  $\psi_{{\rm mI}\alpha}$ can 
similarly be obtained from this $\rho_{{\rm R}}$. 
Single-wall approximation allows us to use the SUSY to obtain 
the bosonic mode functions. 
By using these mode functions we can 
expand the fields 
\begin{equation}
 \psi_{\rm mR}(x^{m},y)=\sum_{n}\rho_{{\rm R}n}(y)\psi_{{\rm mR}n}(x^{m}), 
\end{equation}
for instance, and similarly for other fields. 

Let us apply our improved single-wall approximation in Eq.(\ref{ipsgl}) 
for the Nambu-Goldstone field and to examine the mass splitting of 
massive matter fields. 
If $R$ is large enough, the mode functions for the matter fields on our wall
can be well-approximated by Eq.(\ref{msvmd}). 
In Table~\ref{msv-dm}, we have shown the mass splitting at each mass 
level for a representative case of 
$h/g=3.5$ and $R=10/(\sqrt{g}\Lm)$ and taking the unit of the mass as 
$1/(\sqrt{g}\Lm)$. 
We notice that mass splitting becomes larger for 
 heavier supermultiplets. 
As can be seen from Fig.\ref{msvpfl}, 
heavier modes have wider profile 
in the extra dimension 
and have a larger overlap with the Nambu-Goldstone fermion, which is 
localized on the other wall. 
Since this is likely to be a generic feature of higher massive modes, 
we expect this phenomenon to be generic. 

\begin{table}
\begin{center}
\begin{tabular}{|c||c|c|c|c|c|c|c|} \hline
 $\bar{m^{2}}$ & 0 & 13 & 24 & 33 & 40 & 45 & 48 \\ \hline
 $\Delta \bar{m^{2}}$ & $1.4\times \bar{10^{-7}}$ & $4.2\times \bar{10^{-7}}$ 
 & $5.4\times \bar{10^{-7}}$ & $7.7\times \bar{10^{-7}}$ & 
 $1.3\times \bar{10^{-6}}$ &
 $3.2\times \bar{10^{-6}}$ & $6.5\times \bar{10^{-5}}$ \\ \hline
\end{tabular}
\caption{The mass splitting at each mass level in the case that $h/g=3.5$ 
and $R=10/(\sqrt{g}\Lm)$. Here $m^{2}$ is the squared mass of 
the supermultiplet and $\Delta m^{2}$ is the mass splitting 
between a boson and 
a fermion belonging to the same supermultiplet. 
The unit of the mass is 
$\sqrt{g}\Lm$.} \label{msv-dm}
\end{center}
\end{table}

\begin{figure}
 \leavevmode
 \epsfxsize=12cm
 \epsfysize=4cm
 \centerline{\epsfbox{msvmd.eps}}
 \caption{The profiles of massive modes and the Nambu-Goldstone fermion 
 localized on the other
 wall in the case that  $h/g=3.5$ and $R=10/(\sqrt{g}\Lm)$. 
 (a) is the mode of $m^{2}=13$ and (b) is the mode of $m^{2}=40$.
 The horizontal axis is the coordinate of $y$ in unit of $1/(\sqrt{g}\Lm)$.
 }
 \label{msvpfl}
\end{figure}

\vspace{4mm}

{\large\bf A model with two fields}

\vspace{2mm}

The wall-antiwall system (\ref{acl}) in the previous model 
has a tachyonic mode 
signaling the annihilation of wall-antiwall into vacuum. 
In the following, we will give an example of two walls preserving different halves of 
SUSY which cannot be annihilated into the vacuum. 
The model is a Wess-Zumino model with two chiral superfields
$\Phi$ and $X$, and its superpotential is 
\begin{equation}
 W(\Phi,X)=\frac{m^{2}}{\lm}\Phi-\frac{\lm}{3}\Phi^{3}-\alpha\Phi X^{2},
\end{equation}
where $m$ is a mass parameter and $\lm$ and $\alpha$ are dimensionless 
coupling constants.
This model has four degenerate SUSY vacua and is known to have several 
types of domain wall configurations \cite{shifman,gani}.
Unlike the previous model, we do not compactify the direction of $y$ here. 

BPS domain wall solutions can be obtained from the following 
first-order equations
\begin{equation}
 \frac{{\rm d}\phi}{{\rm d}y}=\Omega\left(\frac{m^{2}}{\lm}-\lm\phi^{2}
 -\alpha\chi^{2}\right), \quad
 \frac{{\rm d}\chi}{{\rm d}y}=\Omega(-2\alpha\phi\chi), \label{bps-eq}
\end{equation}
where $\phi$ and $\chi$ are real parts of 
scalar components of $\Phi$ and $X$ respectively,
and $\Omega$ is a phase factor determined by boundary conditions.
{}For the case of $\alpha=\lm/4$, Eq.(\ref{bps-eq}) has analytic solutions 
representing a wall localized at $y=0$
\begin{equation}
 \Omega=1:\;\phcl^{(1)}(y)=\frac{m}{2\lm}\left(1+\tanh\frac{my}{2}\right),
 \;\;\; \chcl^{(1)}(y)=\frac{\sqrt{2}m}{\lm}\sqrt{1-\tanh\frac{my}{2}},
\end{equation}
which preserves a half of the SUSY called $Q^{(1)}$.
 This will be refered as type I and 
\begin{equation}
 \Omega=-1:\;\phcl^{(2)}(y)=\frac{m}{2\lm}\left(1-\tanh\frac{my}{2}\right),
 \;\;\; \chcl^{(2)}(y)=-\frac{\sqrt{2}m}{\lm}\sqrt{1+\tanh\frac{my}{2}},
\end{equation}
which preserves the other half of the SUSY and will be called type II.

Now we consider a non-BPS configuration constructed from a superposition
of these BPS walls of type I at $y=0$ (our wall) 
and type II at $y=a$ (the other wall) \cite{gani}
\begin{eqnarray}
 \phcl(y)&\!\!\!=&\!\!\!\frac{m}{2\lm}\left(\tanh\frac{my}{2}-\tanh\frac{m(y-a)}{2}
 \right), \nonumber \\
 \chcl(y)&\!\!\!=&\!\!\!\frac{\sqrt{2}m}{\lm}\left(\sqrt{1-\tanh\frac{my}{2}}
 -\sqrt{1+\tanh\frac{m(y-a)}{2}}\right). \label{dbl-wl}
\end{eqnarray}

Note that this is not a static classical solution of the equations of 
motion. 
However, it is approximately a static classical 
solution for large values of $a$, 
which is a distance between two walls. 
In addition, two walls cannot annihilate into a vacuum, 
since the vacua at $y=-\infty$ and $y=\infty$ are different. 
Therefore we have no reason to suspect tachyonic modes. 

We will focus on (nearly) massless modes localized on our wall at $y=0$, 
and evaluate the SUSY breaking mass splitting between them by calculating
the Yukawa coupling in the three-dimensional effective theory.
Integrating over the $y$-direction and integrating out the massive modes,
we obtain a three-dimensional effective  Lagrangian. 
It contains the Yukawa coupling involving 
$a^{(1)}_{0}$ and $\psi^{(1)}_{0}$, which form a $Q^{(1)}$-supermultiplet
and localized on our wall, and the Nambu-Goldstone fermion $\psi_{\rm NG}$ 
associated with the $Q^{(1)}$-SUSY breaking 
\begin{displaymath}
 {\cal L}_{\rm yukawa}^{(3)}=\gef \, a^{(1)}_{0} \psi^{(1)}_{0} \psi_{\rm NG},
\end{displaymath}
\begin{equation}
 \gef=-\frac{\lm}{2\sqrt{2}}\int_{-\infty}^{\infty}{\rm d}y\left(4\rho^{2}\rhng
 +2\rho\sigma\sgng+\sigma^{2}\rhng\right), \label{gani-gef}
\end{equation}
where $\rho$ and $\sigma$ are the $\phi$ and $\chi$-components of the common
mode function of the supermultiplet 
$\left(a^{(1)}_{0}, \psi^{(1)}_{0}\right)$,
and $\rhng$ and $\sgng$ are the $\phi$ and $\chi$-components of 
the Nambu-Goldstone fermion mode function.
Using the single-wall approximation, we obtain mode functions for the 
supermultiplet $\rho, \sigma$ and for the Nambu-Goldstone fermion 
$\rho_{\rm NG}, \sigma_{\rm NG}$ 
\begin{eqnarray}
 \rho(y)=\frac{C}{\cosh^{2}\frac{my}{2}}, \quad
 \sigma(y)= -\sqrt{2}C\left(1+\tanh\frac{my}{2}\right)
 \sqrt{1-\tanh\frac{my}{2}},
\end{eqnarray}
\begin{eqnarray}
 \rhng(y)&\!\!\!=&\!\!\!\frac{C}{\cosh^{2}\frac{m(y-a)}{2}}, \nonumber \\
 \sgng(y)&\!\!\!=&\!\!\! -\sqrt{2}C\left(1+\tanh\frac{m(y-a)}{2}\right)
 \sqrt{1-\tanh\frac{m(y-a)}{2}},
\end{eqnarray}
where $C$ is a normalization factor and is taken to be positive.

Substituting these mode functions into Eq.(\ref{gani-gef}), 
we obtain the value of the effective Yukawa coupling $\gef \approx 0.520 \lambda \sqrt{m} e^{-0.50 am}$ for $ am \gg 1$ 
. 
{}From the energy density of the background field configuration 
we obtain also the SUSY breaking order parameter 
 $f \approx 2\sqrt{2/3} (m^{3/2}/\lambda)$. 
Therefore we find that the mass splitting between boson and fermion 
as $\Delta m^2 = m_{\rm B}^2 - m_{\rm F}^2 \approx 1.20 m^2 e^{-0.50 am}$. 
The fact that the mass squared of the bosonic mode shifts to 
positive values\footnote{The fermionic zero mode remains massless due to 
its property of the Nambu-Goldstone fermion.} 
is in accord with our observation of the absence of tachyon 
corresponding to the annihilation into the vacuum.

\begin{center}
{\bf Acknowledgements}
\end{center}
This work is supported in part by Grant-in-Aid for Scientific 
Research from the Ministry of Education, Science and Culture 
for the Priority Area 291 and 707. 
N.M.,Y.S. and R.S. are supported 
by the Japan Society for the Promotion of Science for Young Scientists 
(No.08557, No.10113 and No.6665).

\end{document}